\documentstyle[prb,aps,multicol]{revtex}
  \renewcommand{\narrowtext}{\begin{multicols}{2} \global\columnwidth20.5pc}
  \renewcommand{\widetext}{\end{multicols} \global\columnwidth42.5pc}

 \input{psfig}

\begin{document}

\draft



\preprint{CondMat}

\title{AC Conductance in Dense Array of the Ge$_{0.7}$Si$_{0.3}$ Quantum Dots in Si}

\author{I. L. Drichko, A. M. Diakonov, I. Yu. Smirnov}

\address{A. F. Ioffe  Physico-Technical Institute of Russian
  Academy of Sciences, 194021
  St. Petersburg, Russia}

\author{A. V. Suslov}

\address{National High Magnetic Field Laboratory, Tallahassee, Florida 32310, USA}

\author{Y. M. Galperin}

\address{Department of Physics and Center for Advanced Materials, University of Oslo, 0316 Oslo, Norway}
\address{A. F. Ioffe  Physico-Technical Institute of Russian Academy of Sciences, 194021
  St. Petersburg, Russia}

\author{A. I. Yakimov and A. I. Nikiforov}

\address{Institute of Semiconductor Physics, Siberian division
of Russian Academy of Sciences, Novosibirsk, Russia}

\date{\today}

\maketitle

\begin{abstract}

Complex AC-conductance, $\sigma^{AC}$, in the systems with dense
Ge$_{0.7}$Si$_{0.3}$ quantum dot (QD) arrays in Si has been
determined from simultaneous measurements of attenuation,
$\Delta\Gamma=\Gamma(H)-\Gamma(0)$, and velocity, $\Delta V
/V=(V(H)-V(0)) / V(0)$, of surface acoustic waves (SAW) with
frequencies $f$ = 30-300 MHz as functions of transverse magnetic
field $H \leq$ 18 T in the temperature range $T$ = 1-20 K.  It has
been shown that in the sample with dopant (B) concentration 8.2$
\times 10^{11}$ cm$^{-2}$ at temperatures $T \leq$4 K the AC
conductivity is dominated by hopping between states localized in
different QDs. The observed power-law temperature dependence,
$\sigma_1(H=0)\propto T^{2.4}$, and weak frequency dependence,
$\sigma_1(H=0)\propto \omega^0$, of the AC conductivity are
consistent with predictions of the two-site model for AC hopping
conductivity for the case of $\omega \tau_0 \gg $1, where
$\omega=2\pi f$ is the SAW angular frequency and $\tau_0$ is the
typical population relaxation time. At $T >$ 7 K the AC
conductivity is due to thermal activation of the carriers (holes)
to the mobility edge. In intermediate temperature region 4$ < T<$
7 K, where AC conductivity is due to a combination of hops between
QDs and diffusion on the mobility edge, one succeeded to separate
both contributions. Temperature dependence of hopping contribution
to the conductivity above $T^*\sim$ 4.5 K saturates, evidencing
crossover to the regime where $\omega \tau_0 < $1. From crossover
condition, $\omega \tau_0(T^*)$ = 1, the typical value, $\tau_0$,
of the relaxation time has been determined.

\end{abstract}
\pacs{73.21.La, 77.65Dg, 72.20.Ee, 72.20.My, 72.50.+b}

\narrowtext

\section{Introduction}
\label{Introduction}

Understanding of AC conductivity of low dimensional semiconductor
structures is important both for fundamental science and
nanodevice applications.  Though the surface acoustic waves (SAW)
technique has been successfully applied for AC conductivity
measurements in two dimensional systems, only limited number of
works aim at zero-dimensional objects \cite{1,2}.  Here we report
our results on the interaction between arrays of
Ge$_{0.7}$Si$_{0.3}$ dots in Si with surface acoustic waves.

\begin{figure}[t]
\centerline{\psfig{figure=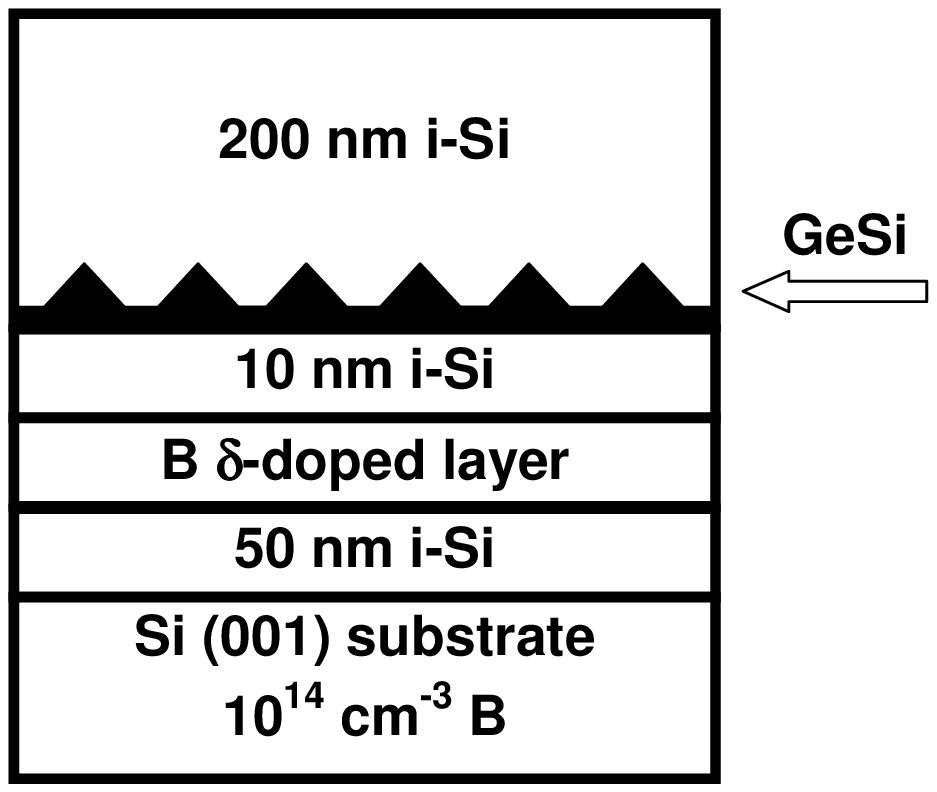,width=5.5cm,clip=} }
 \caption{Layer scheme of the sample containing a dense
array of Ge$_{0.7}$Si$_{0.3}$ quantum dots. \label{Sample}}
\end{figure}

A layer containing a dense (3 $\times 10^{11}$ cm$^{-2}$)
self-assembled array of Ge$_{0.7}$Si$_{0.3}$ quantum dots has been
grown on a B delta-doped Si substrate and then covered by a 2000
$\AA$ Si layer (Fig.~\ref{Sample}).  The dots were square pyramids
with 120$\times$120 $\AA^2$ in base and 12 $\AA$ in height.  Three
samples with distinct B concentrations - 6.8 $\times 10^{11}$
cm$^{-2}$ (sample 1), 8.2$\times 10^{11}$ cm$^{-2}$ (sample 2),
and 11 $\times 10^{11}$ cm$^{-2}$ (sample 3) - were investigated.

\begin{figure}[h]
\centerline{\psfig{figure=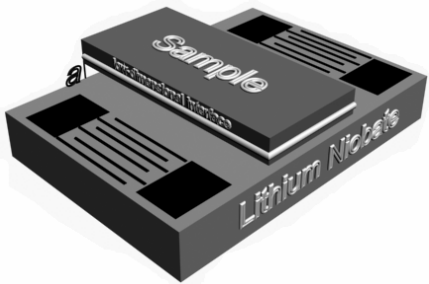,width=6cm,clip=} }
 \caption{Scheme of the acoustoelectric device. The electric field
of a surface acoustic wave propagating on the surface of a
piezoelectric substrate acts on a low-dimensional electron/hole
system "embedded" into the sample close to its surface. This
"hybrid" geometry allows applying a sliding electrostatic
potential to the electron/hole system in non-piezoelectric
materials. \label{SetUp}}
\end{figure}

We studied the influence of an external magnetic field on the
attenuation and relative velocity change for a SAW propagating
along a surface of a piezoelectric lithium niobate platelet, when
a sample with the quantum dots array was pressed to the surface
(Fig.~\ref{SetUp}). The AC electric field accompanying the SAW
penetrated into the sample, while no mechanical strain in the
semiconductor sample was produced.  The applied magnetic field of
up to 18 T was perpendicular to the quantum dots layer, and
therefore to the SAW wave vector.  The measurements were performed
in the temperature range 1-20 K and in the SAW frequency range
from 30 to 300 MHz.

\section{Experimental Results and Discussion}

\label{Experiment}

\begin{figure}[h]
\centerline{\psfig{figure=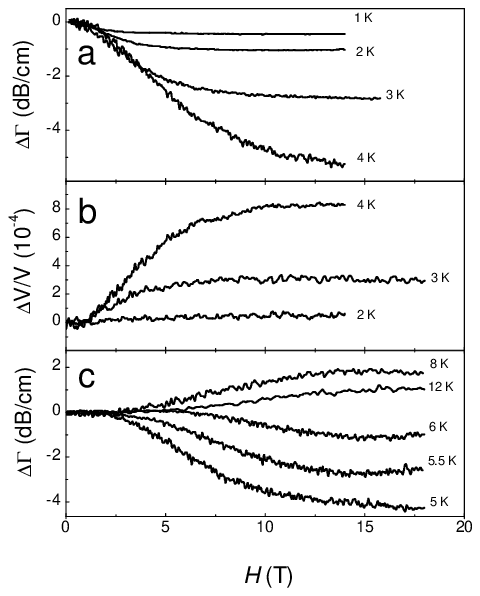,width=6.5cm,clip=} }
 \caption{Magnetic field dependence of attenuation $\Delta\Gamma$
 (a), (c) and velocity $\Delta V/V$ (b)
 at different temperatures; sample 2; $f$ = 28 MHz. \label{GVH}}
\end{figure}

In the temperature interval between 1 and 5 K, see Fig.~\ref{GVH}
(a-b), the SAW attenuation decreases, while the SAW velocity
increases with magnetic field. We ascribe these behaviors to the
inter-dot AC hopping conductance, described by the conventional
two-site model \cite{3}. According to this model, variations of
the SAW attenuation and velocity with magnetic field is due to
shrinking of the localized wave functions of the states involved
into hopping.  The theory \cite{3} predicts that in low magnetic
fields both $\Delta\Gamma=\Gamma(H)-\Gamma(0)$ and $\Delta V
/V=(V(H)-V(0)) / V(0)$ are proportional to $H^2$, while at high
fields both quantities are proportional to $H^{-2}$, and therefore
both the attenuation and velocity saturate. Our experimental data
agree with these predictions.

\begin{figure}[h]
\centerline{\psfig{figure=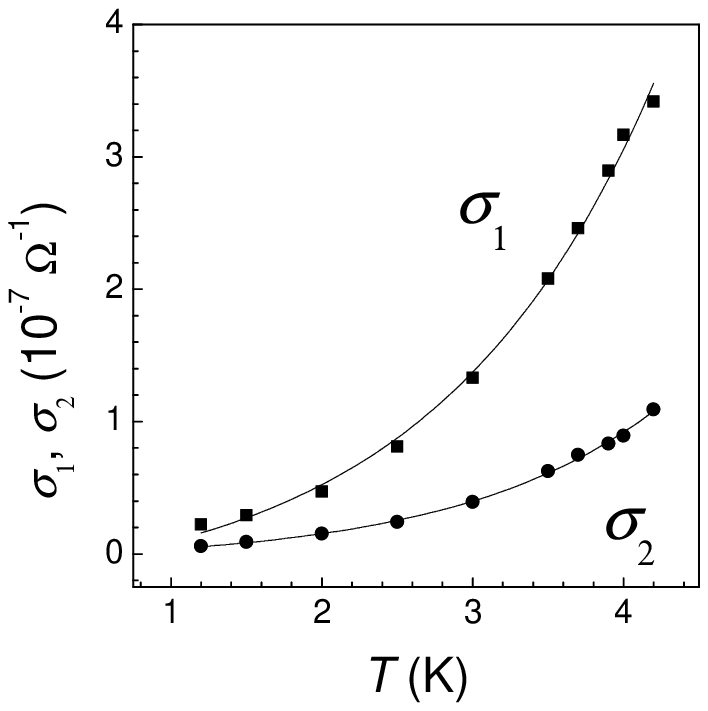,width=5cm,clip=} }
 \caption{Dependences of the real $\sigma_1 (H=0)$ and imaginary $\sigma_2 (H=0)$
 components of the AC conductivity on a temperature; sample 2; $f$ = 28 MHz. \label{s12}}
\end{figure}

Moreover, since $\Delta\Gamma \mid _{H \rightarrow
\infty}=-\Gamma(0)+A/H^2$ one can find the zero-field attenuation,
$\Gamma(0)$. A similar procedure can be used to determine  $\Delta
V(0)/V$. Thus, simultaneous measurement of $\Delta\Gamma (H)$ and
$\Delta V(H)$ allows us to implement the procedure described in
Refs.~\onlinecite{prbild,prbsmirnov} for determining the zero
field AC conductance, $\sigma=\sigma_1-i\sigma_2$, Fig.~\ref{s12},
for different temperatures and SAW frequencies. It has been found
that both $\sigma_1$ and $\sigma_2$ are frequency-independent with
accuracy 15$\%$ and 25$\%$, respectively.  Power-law temperature
dependence of conductance, $\sigma_1 (H=0)\propto T^{2.4}$, and
its independence of frequency agree with predictions of the
two-site model for the case $\omega \tau_0 \gg 1$, where $\tau_0$
is the population relaxation time for a typical pair of dots
giving contribution to the conductance.

At higher temperatures the sign of $\Delta\Gamma$ changes from
negative to positive. At the same time, the two-site model
predicts the crossover to the regime $\omega \tau_0 \ll$1, where
$\Delta\Gamma$ should saturate as a function of the temperature.
To explain the experimental temperature dependence we assume that
with temperature increase an additional conduction mechanism -
carrier activation to the mobility edge - emerges and becomes
dominant. As it follows from the experimental data, at higher
temperatures the real part of the conductivity follows the
activation law, $\sigma_1 (H=0) \propto \exp (-E_a/k_B T)$, with
$E_a$=2.5 meV.

\begin{figure}[h]
\centerline{\psfig{figure=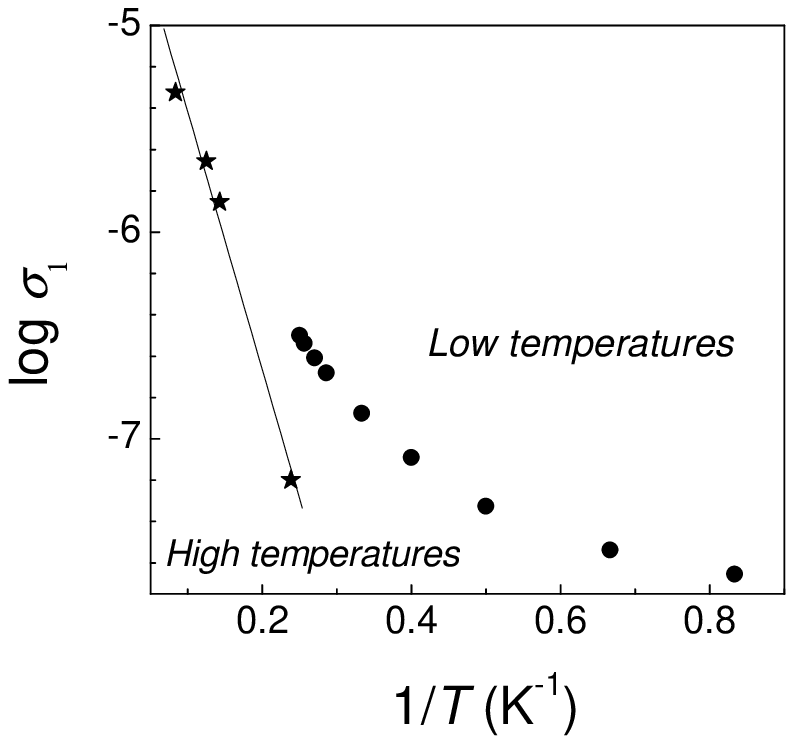,width=5.5cm,clip=} }
 \caption{Real part of the AC conductivity $\sigma_1 (H=0)$ vs. 1/$T$; sample 2; $f$ = 28 MHz.
 The line shows the contribution
 from the extended states. \label{logs1}}
\end{figure}

\begin{figure}[h]
\centerline{\psfig{figure=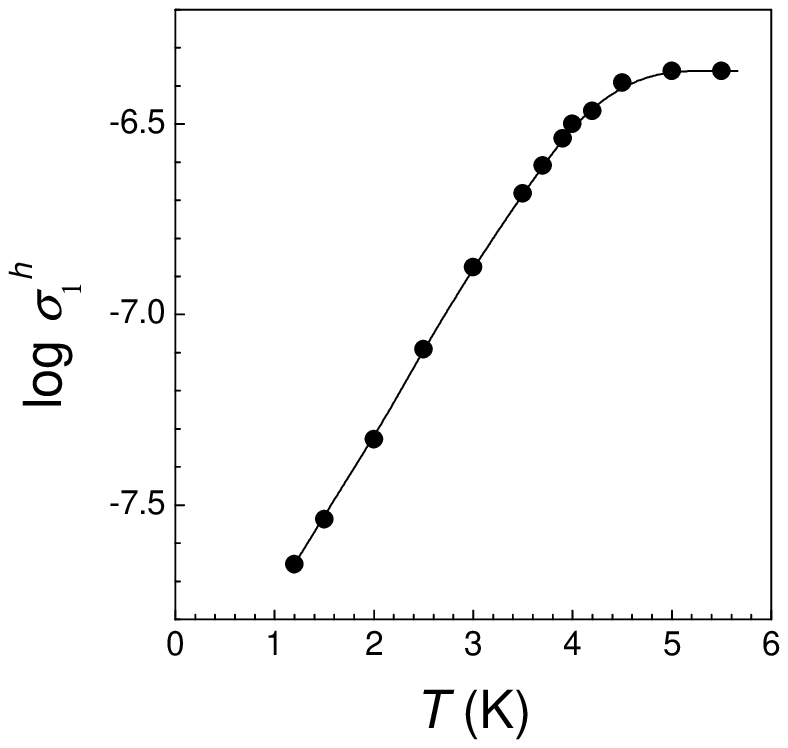,width=5.5cm,clip=} }
 \caption{Temperature dependence of the hopping contribution to the conductivity $\sigma_1^h (H=0)$;
 sample 2; $f$ = 28 MHz.
  \label{s1h}}
\end{figure}

At intermediate temperatures, 4$<T<$7 K, the AC conductivity in
the sample 2 is a combination of hops between quantum dots and
diffusion at the mobility edge.  Extrapolation of the activation
dependence to lower temperatures, Fig.~\ref{logs1}, allowed us to
extract the hopping contribution,  $\sigma_1^h$, in the whole
temperature range, Fig.~\ref{s1h}. It is seen that  $\sigma_1^h$
saturates as a function of the temperature at temperatures higher
than $T^* \sim$ 4.5 K, evidencing transition to the regime where
$\omega \tau_0<$1. It follows from the crossover condition,
$\omega \tau_0 (T^*)$=1, that $\tau_0 (T^*)\approx 5\times
10^{-9}$s.

It is worth noting that the two-site model predicts $\sigma_2$ to
be greater than $\sigma_1$, however, that is not the case in the
experiment. We hope that a more sophisticated model allowing for
the properties of dense arrays will provide better quantitative
description of the experimental situation.

\acknowledgments This work is supported by RFFI 04-02-16246,
03-02-16526 and Presidium RAN grants. Part of the work was
performed at the NHMFL, Tallahassee, FL, which is operated under
patronage of the NSF (DMR-0084173) and State of Florida. The
ultrasonic research at the NHMFL is supported by an IHRP grant.


\widetext

\end{document}